\documentclass[aps,prl,twocolumn,showpacs,groupedaddress,apsrev,revsymb]{revtex4}
\usepackage{amssymb}
\usepackage{amsmath}
\usepackage{graphicx}
\usepackage{amsbsy}
\usepackage{revsymb}
\usepackage{dcolumn}
\usepackage{bm}

\def\C{{\@QC C}}
\def\@QC#1{\mathpalette{\setbox0=\hbox\bgroup$\rm}  {\egroup C$\egroup\rm\rlap{\kern0.4\wd0\vrule
  width 0.05\wd0 height 0.97\ht0 depth -0.01\ht0}  #1\bgroup}}

\begin{document}

\title{An Efficient and Accurate Car-Parrinello-like \\
Approach to Born-Oppenheimer Molecular Dynamics}
\author{Thomas D. K\"uhne}
\email{tkuehne@phys.chem.ethz.ch}
\affiliation{Computational Science, Department of Chemistry and Applied Biosciences, ETH
Zurich, USI Campus, Via Giuseppe Buffi 13, CH-6900 Lugano, Switzerland}
\author{Matthias Krack}
\affiliation{Computational Science, Department of Chemistry and Applied Biosciences, ETH
Zurich, USI Campus, Via Giuseppe Buffi 13, CH-6900 Lugano, Switzerland}
\author{Fawzi R. Mohamed}
\affiliation{Computational Science, Department of Chemistry and Applied Biosciences, ETH
Zurich, USI Campus, Via Giuseppe Buffi 13, CH-6900 Lugano, Switzerland}
\altaffiliation{Scuola Normale Superiore di Pisa, Piazza dei Cavalieri 7, I-56125 Pisa, Italy}
\author{Michele Parrinello}
\affiliation{Computational Science, Department of Chemistry and Applied Biosciences, ETH
Zurich, USI Campus, Via Giuseppe Buffi 13, CH-6900 Lugano, Switzerland}
\altaffiliation{Scuola Normale Superiore di Pisa, Piazza dei Cavalieri 7, I-56125 Pisa, Italy}
\date{\today }

\begin{abstract}
We present a new method which combines Car-Parrinello and Born-Oppenheimer molecular dynamics in order to accelerate density functional theory based ab-initio simulations. Depending on the system a gain in efficiency of one to two orders of magnitude has been observed, which allows ab-initio molecular dynamics of much larger time and length scales than previously thought feasible. It will be demonstrated that the dynamics is correctly reproduced and that high accuracy can be maintained throughout for systems ranging from insulators to semiconductors and even to metals in condensed phases. This development considerably extends the scope of ab-initio simulations.
\end{abstract}

\maketitle

Density functional theory (DFT) \cite{KohnSham1965} based ab-initio molecular dynamics \cite{CarParrinello1985}, in which interactions are computed on the fly from electronic structure calculations, has been very successful in describing a large variety of physical phenomena and has proven its relevance in many fields. However, its computational cost has limited the attainable length and time scales in spite of substantial progress. For a while it was believed that the development of linear scaling methods \cite{Yang1991, GalliParrinello1992, Goedecker1999} could have offered a solution. Unfortunately, the crossover point at which linear scaling methods become advantageous has remained fairly large, especially if high accuracy is needed.

On the other hand, it would be very desirable to accelerate ab-initio simulations with up to thousands of atoms, such that simulations as long as tens or even hundreds of picoseconds can be routinely performed, thus making completely new phenomena accessible to ab-initio simulations. Born-Oppenheimer molecular dynamics (BOMD) simulations, in which at each time step the DFT functional is fully minimized, do not seem to offer much room for further improvement. For this reason another direction has recently been followed to improve the efficiency at current system sizes. In the spirit of Car-Parrinello molecular dynamics (CPMD) \cite{CarParrinello1985}, some form of dynamics for the electronic degrees of freedom is implemented, which automatically maintains the system close to the BO surface, but at variance with the original proposal in a localized orbital representation \cite{Iyengar2001,SharmaCar2003, ThomasTuckerman2004, HerbertHeadGordon2004}. The acceleration stems from the ability to reduce or fully bypass the self-consistency cycle. However, just like in CPMD, these methods suffer from rather short integration time steps. In this Letter we present a new method
which overcomes this limitation and combines the accuracy and long time steps of BOMD with the efficiency of CPMD.

We consider the general case in which the DFT Kohn-Sham orbitals are expanded in a non-orthogonal basis. Let $M$ be the dimension of the Hilbert space and $\mathbf{S}$ the $M\times M$ overlap matrix. As is customary in quantum chemistry we arrange the expansion coefficients of the $N$ lowest occupied orbitals in a rectangular $M\times N$ matrix $\mathbf{C}$. The density matrix is then written as $\mathbf{P}=\mathbf{CC}^{T}$ and obeys the idempotency condition $\mathbf{P}=\mathbf{PSP}$. Maintaining the idempotency of $\mathbf{P}$ is crucial in any electronic structure calculation and is one of its algorithmic challenges. The potential energy surface on which the ions move is defined by the minimum of an appropriately chosen energy functional $E_{\text{DFT}}\left[\mathbf{C},\mathbf{R}_{I}\right]$, which we express as a functional of $\mathbf{C}$ and a function of the ionic coordinates $\mathbf{R}_{I}$. In
this notation the Born-Oppenheimer equation of motions reads as follows:%
\begin{equation}
M_{I}{\mathbf{\Ddot{R}}}_{I}=-\nabla _{I}\min_{\mathbf{C}}E_{\text{DFT}}\left[\mathbf{%
C},\mathbf{R}_{I} \right] \mbox{,}  \label{BOMD_EOM}
\end{equation}%
where the search for the minimum is restricted to the $\mathbf{C}$'s that satisfy the orthonormality condition $\mathbf{C}^{T}\mathbf{SC=I}$, which is equivalent to imposing the idempotency condition on $\mathbf{P}$. The forces of Eq.~(\ref{BOMD_EOM}) can be divided into three contributions, the
Hellmann-Feynman forces \cite{Hellmann1933, Feynman1939}, the Pulay forces \cite{Pulay1969}, which are present whenever the basis set depends on the ionic positions, and a residual term \cite{BendtZunger1983} which is non-zero except when full self-consistency is reached. This term leads to poor energy conservation in BOMD unless a very tight convergence criterion is imposed. In Car-Parrinello-like approaches this is circumvented by the design of a coupled electron-ion dynamics which maintains the system close to the BO surface, but at the cost of small time steps.

Based on ideas of the original Car-Parrinello approach we design an improved dynamics for the coupled system of electrons and ions. Contrary to the Car-Parrinello scheme we will not write an explicit equation of motion for the $\mathbf{C}$'s, but rather an integration scheme for the electronic degrees of freedom. The knowledge of the previous $K$ values of $\mathbf{C}\left( t_{n-l}\right)$, where $l \in [1,K]$, determines the value of $\mathbf{C}\left( t_{n}\right)$, such that at any instant of time the $\mathbf{C}$'s are as close as possible to the instantaneous ground state. As for the short-term integration of the electronic degrees of freedom, accuracy is crucial, a highly accurate and efficient algorithm is required. Here we have chosen the always stable predictor-corrector (ASPC) method \cite{Kolafa2004} of Kolafa. This method was originally designed to deal with classical polarization, so that additional care must be taken that during the evolution the idempotency condition is satisfied. The predictor
\begin{eqnarray}
\mathbf{C}^{p}\left( t_{n}\right) &\cong& \sum_{m=1}^{K}(-1)^{m+1}m%
\frac{{\binom{2K}{K-m}}}{{\binom{2K-2}{K-1}}}\underbrace{\mathbf{C}(t_{n-m})%
\mathbf{C}^{T}(t_{n-m})}_{\mathbf{P}(t_{n-m})} \nonumber \\
&&\times \mathbf{S}(t_{n-m}) \mathbf{C}\left( t_{n-1}\right)
\label{ASPC}
\end{eqnarray}%
uses the extrapolated contra-covariant density matrix $\mathbf{PS}$ \cite{CurvySteps2003} as an 
approximate projector on to the occupied subspace $\mathbf{C} \left( t_{n-1}\right)$. In this way, we take advantage of the fact that the physically relevant contra-covariant density matrix $\mathbf{PS}$ evolves
more smoothly and is therefore easier to predict than $\mathbf{C}$. This is followed by a corrector step to minimize the error and to further reduce the deviation from the ground state. The corrector
\begin{eqnarray}
\mathbf{C}\left( t_{n}\right)  &=&\omega \, \text{MIN}\left[\mathbf{C}^{p}
\left( t_{n}\right)\right] +(1-\omega )\mathbf{C}^{p}\left( t_{n}\right), \nonumber \\
\mbox{with~}\omega  &=&\frac{K}{2K-1} %\nonumber
\end{eqnarray}%
consists of a single minimization step $\text{MIN}\left[\mathbf{C}^{p}(t_{n})\right]$ 
of a properly selected minimization procedure. Alternatively, the predictor can also be repeatedly applied. In this case the ground state is even more closely approached, but in general this is not necessary. The numerical coefficients of Eq.~(\ref{ASPC}) were selected by Kolafa in order to ensure time-reversibility up to $O(h^{K+2})$, while $\omega$ was chosen to guarantee a stable relaxation towards the minimum. Because the energy is invariant under a unitary transformation within the subspace of occupied orbitals $\mathbf{C}$, it must be ensured that this gauge transformation is not strongly changed by $\text{MIN}\left[ \mathbf{C}^{p}\left(t_{n}\right) \right]$, as in this case continuity between the $\mathbf{C}$'s may be lost. Moreover the minimization scheme must be very efficient in bringing the system close to the ground state and preserving the idempotency of the density matrix. For these reasons we have chosen the orbital transformation (OT) method of VandeVondele and Hutter 
\cite{OT2003}. Inspired by the form of the exponential transformation \cite{HutterParrinello1994} an auxiliary variable $\mathbf{X}$ is introduced, to parameterize the occupied orbitals
\begin{subequations}
\begin{equation}
\mathbf{C} \left( \mathbf{X}\right) =\mathbf{C}^{p}\left( t_{n}\right) \cos
\left( \mathbf{U}\right) +\mathbf{X}\mathbf{U}^{-1}\sin \left( \mathbf{U}%
\right) 
\end{equation}%
where  $\mathbf{U=}\left( \mathbf{X}^{T}\mathbf{SX}\right) ^{1/2}$ and the
variable $\mathbf{X}$ has to obey the linear constraint:%
\begin{equation}
\mathbf{X}^{T}\mathbf{SC}^{p}\left( t_{n}\right) =0
\label{constraint}
\end{equation}%
\end{subequations}
Under this condition $\mathbf{C} \left(\mathbf{X}\right)$ leads to an idempotent density matrix for any choice of $\mathbf{X}$, provided that the reference orbitals $\mathbf{C}^{p}\left( t_{n}\right)$ are orthonormal. Thus a finite minimization step along the gradient direction will still fulfill the idempotency constraint exactly. Due to the linear constraint the minimization with respect to $\mathbf{X}$ is performed in an auxiliary tangent space. Because this is a linear one no curved geodesics must be followed, as is the case for variables such as $\mathbf{C}$, which are nonlinearly constrained. In this way large minimization steps can be taken, especially if a good preconditioner is used. In fact using a very efficient, idempotency conserving direct minimizer like OT has been decisive for the success of this approach. Since the ASPC integrator preserves the orthonormality constraint only approximately, it occasionally has to be explicitly enforced, either by a Cholesky decomposition or by a few purification iterations \cite{McWeeny1960}.

Having obtained the new wavefunction we can now evaluate the energy and the nuclear forces, which are derived from the following approximate energy functional:
\begin{eqnarray}
E_{\text{PC}}[\rho ] &=&{\text{Tr}}\left[ \mathbf{C}^{T}H[\rho
^{p}] \, \mathbf{C}\right] -\frac{1}{2}\int {d\mathbf{r}\int {d\mathbf{r^{\prime} }}%
\, \frac{\rho ^{p}(\mathbf{r})\rho ^{p}(\mathbf{r}^{\prime })}{|\mathbf{r}-%
\mathbf{r}^{\prime }|}}  \notag \\
&-&\int {d\mathbf{r} \,V_{\text{XC}}[\rho }^{p}{]}\rho ^{p}+E_{\text{XC}}[{\rho }%
^{p}]+E_{II} \mbox{,}
\label{DensityFunctional}
\end{eqnarray}%
where $\rho ^{p}(\mathbf{r})$ is the density associated with $\mathbf{C}^{p}(t_{n})$. $E_{\text{PC}}[\rho]$ can be thought of as an approximation to the Harris-Foulkes functional \cite{HarrisFoulkes} and maintains the predictor-corrector flavor of the present work. The validity of $E_{\text{PC}}[\rho]$ depends only on the efficiency of the minimizer and on the quality of the propagation scheme. The ionic forces are calculated by evaluating the analytic gradient of $E_{\text{PC}}[\rho]$, with respect to the nuclear coordinates. However, as  $\Delta{\rho} = \rho - \rho^{p} \neq 0$, besides the usual Hellmann-Feynman and Pulay forces an extra term appears: 
\begin{equation}
-\int {d\mathbf{r}\left\{ \left[ \left( \frac{\partial V_{\text{XC}}[{\rho}^{p}]%
}{\partial {\rho}^{p}}\right) \Delta {\rho}+V_{\text{H}}[\Delta {\rho}]%
\right] \left(\nabla _{I}{\rho }^{p} \right) \right\} } \mbox{,}
\label{FNV}
\end{equation}%
where $\rho$ is the corrected density evaluated using $\mathbf{C}(t_{n})$ and ${\rho}^{p}$ is the predicted density calculated from $\mathbf{C}^{p}(t_{n})$. However, as a single minimization step is performed, $\mathbf{C}(t_{n})$ is only an approximate eigenfunction of $H[\rho^{p}]$ within the subspace spanned by the finite basis set used \cite{MarxHutter2000}. This leads to an insignificant error in the forces, provided that $\mathbf{C}(t_n)$ is very close to the ground-state.

The ability of this dynamics to maintain the system on the BO surface may vary considerably. It is almost perfect in systems like water, but somewhat less satisfactory in liquid Si at high temperature, where swift bonding and rebonding processes take place. However, in all cases the dynamics is dissipative, most likely due to the fact that the propagation scheme employed is not symplectic. It is possible to remedy this downward drift if we assume that the forces arising from our dynamics $\mathbf{F}_{\text{PC}}$ can be modeled as $\mathbf{F}_{\text{PC}}=\mathbf{F}_{\text{BO}}-\gamma_{D}{\mathbf{\Dot{R}}}_{I}$ which, as we shall see later, is an excellent assumption. The value of the intrinsic friction coefficient $\gamma_{D}$ does not need to be known but it can be bootstrapped by taking a cue from the work of Krajewski and Parrinello \cite{KrajewskiParrinello2006}. We sample the canonical distribution by using the following Langevin-type equation%
\begin{equation}
M_{I}{\mathbf{\Ddot{R}}}_{I}=\mathbf{F}_{\text{PC}}-\gamma_{L}{\mathbf{\Dot{R}}}%
_{I}+\mathbf{\Xi }_{I} \mbox{,}
\label{LangevinEq}
\end{equation}%
where $M_{I}$ is the ionic mass, $\gamma_{L}$ is a Langevin friction coefficient and $%
\mathbf{\Xi }_{I}=\mathbf{\Xi}_{I}^{D}+\mathbf{\Xi}_{I}^{L}$ a random noise. 
Using the assumption above this can be equally written as:
\begin{equation}
M_{I}{\mathbf{\Ddot{R}}}_{I}=\mathbf{F}_{\text{BO}}-\left( \gamma_{D}+\gamma
_{L}\right) {\mathbf{\Dot{R}}}_{I}+\mathbf{\Xi }_{I} %\mbox{.}
\label{revLangevinEq}
\end{equation}%
In order to guarantee an accurate sampling of the Boltzmann distribution, the noise has to obey the fluctuation dissipation theorem:%
\begin{equation}
\left\langle \mathbf{\Xi }_{I}\left( 0\right) \mathbf{\Xi }_{I}\left(
t\right) \right\rangle =6\left( \gamma_{D}+\gamma_{L}\right)
M_{I}k_{B}T\delta \left( t\right) %\mbox{,}
\end{equation}%
The choice of $\gamma_{L}$ is arbitrary, while the unknown $\gamma_{D}$ has to be determined by requiring that the aggregate noise term generate the correct average temperature, as measured by the equipartition theorem $\left\langle \frac{1}{2}M_{I}{\mathbf{\Dot{R}}}_{I}^{2}\right\rangle =\frac{3}{2}k_{\text{B}}T$. We will show that this leads to correct sampling. Since our initial dynamics is quite accurate $\gamma_{D}$ is small, so that dynamical properties are also well reproduced.

For the purpose of demonstrating our new method, we have implemented it in the mixed Gaussian Plane Wave (GPW) \cite{Lippert1997} code \textsc{Quickstep} \cite{VandeVondele2005} which is part
of the publicly available suite of programs CP2K \cite{CP2K}. We present here calculations on metallic liquid silicon and liquid silica, to illustrate that our method works well irrespective of band-gap, system size and system type. Both systems are known to be very difficult, and are examples of liquid metals ($\text{Si)}$ and of complex, highly polarizable, ionic liquids ($\text{SiO}_{\text{2}})$. In addition the simulations have been performed at 3000~K and 3500~K respectively, which leads to rapidly varying density matrix elements, thus making the propagation of the electronic degrees of freedom quite challenging. Hence, the selected tests can be considered as worst-case scenarios for our method.

All simulations have been performed at their experimental liquid densities using double-zeta valence polarization (DZVP) basis sets, adequate density cutoffs, the Goedecker-Teter-Hutter pseudopotentials \cite{GTH1996} and the local-density approximation to the exact exchange and correlation functional. 
For simplicity the Brillouin zone is sampled at the $\Gamma $-point only. Eq.~(\ref{revLangevinEq}) is integrated using the algorithm of Ricci and Ciccotti \cite{RicciCiccotti2003}, with a time step of $h=1.0~\text{fs}$. The friction coefficient $\gamma_{L}$ was set equal to zero, while the values for  $\gamma_{D}$ turned out to be in the range of $10^{-4}~\text{fs}^{-1}$. We predict the new $\mathbf{C}$'s using $K=4$ in Eq.~(\ref{ASPC}), which ensures time-reversibility up to $O(h^{6})$. The OT minimizer is preconditioned, as proposed in \cite{OT2003}. 

\begin{figure}[tbp]
\includegraphics[width=8.6cm]{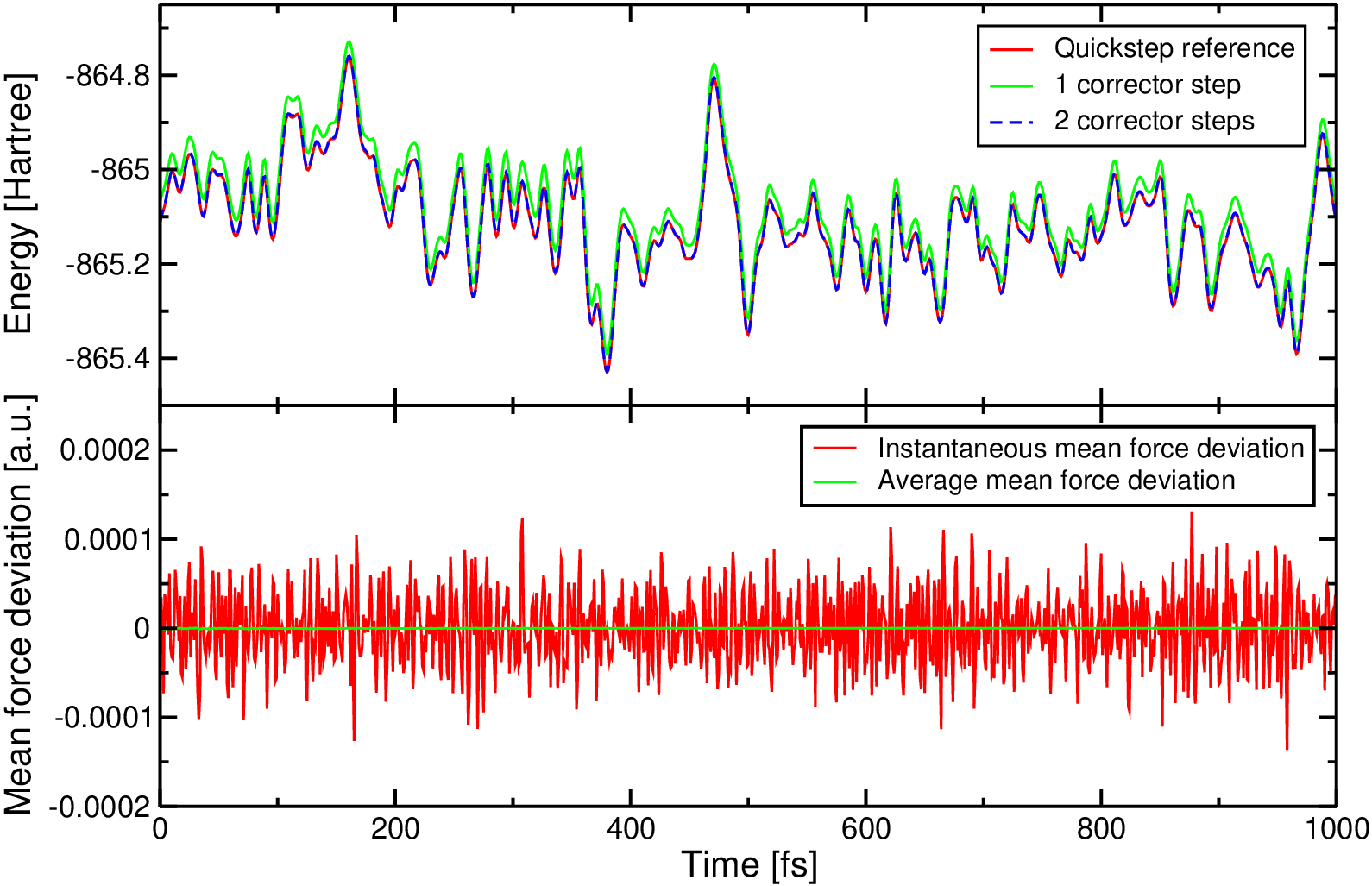}
\caption{Deviations from the BO surface of liquid $\text{SiO}_{\text{2}}$ with respect to total energies (upper panel) and mean force deviations (lower panel). The deviation in the energies corresponds to a constant shift of $4.16\times10^{-4}$ Hartree per atom for one corrector step and $3.5\times10^{-5}$ Hartree per atom for two corrector steps. The average mean force deviation is unbiased.} 
\label{RefTraj}
\end{figure}

We first consider the accuracy in terms of deviation from the BO surface. As can be seen in FIG.~\ref{RefTraj} our energies are an upper bound to the ground state and are displaced by a small and approximately constant amount. It is also shown that the deviation from the BO surface can be even further reduced by increasing the number of corrector steps. In fact the deviation can be controlled by varying the number of corrector steps in order to achieve a preassigned accuracy level. In the following only simulations based on a single corrector step will be reported.

\begin{figure}[tbp]
\includegraphics[width=8.6cm]{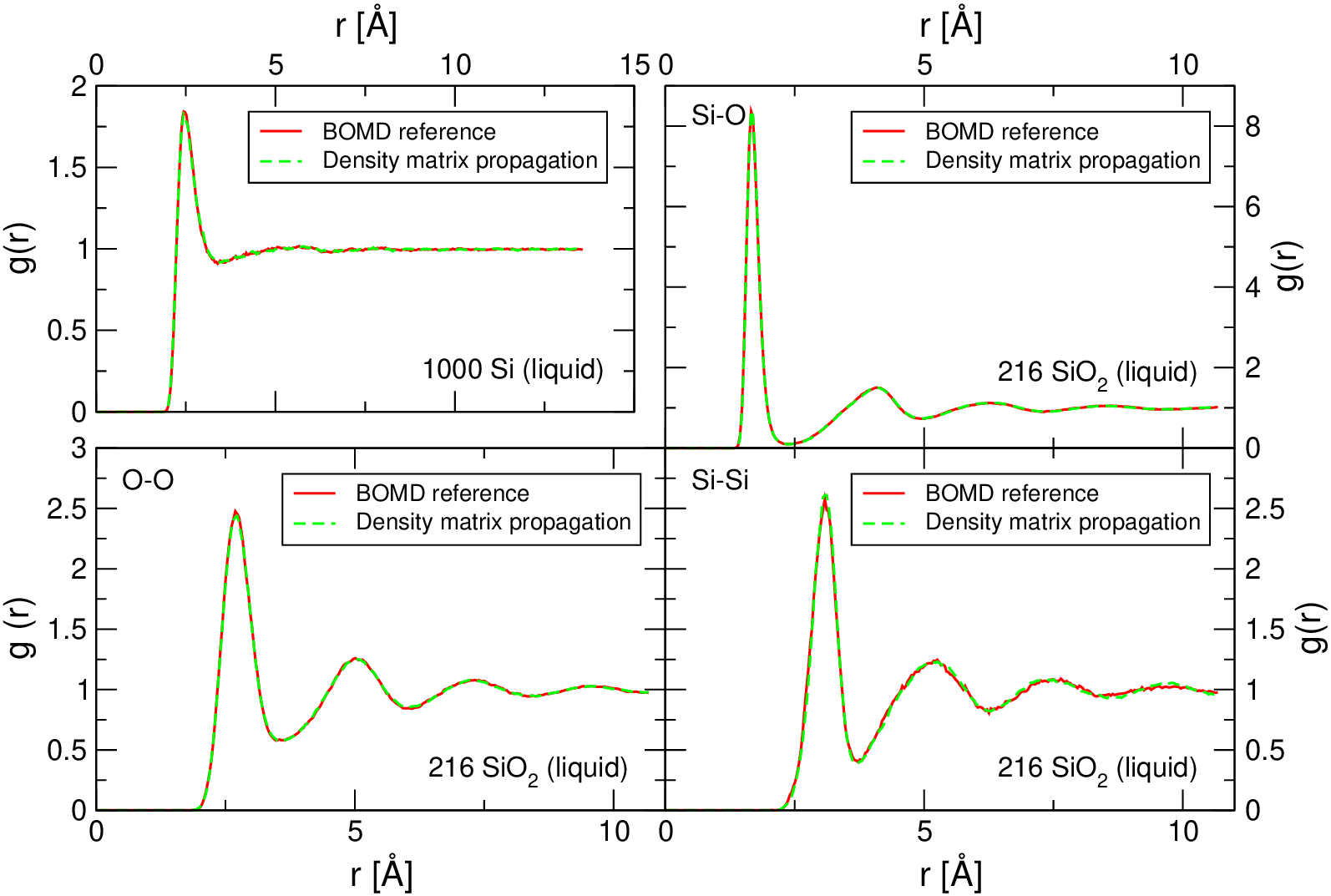}
\caption{Partial pair-correlation functions g(r) of liquid $\text{Si}$ (upper left panel) and liquid $\text{SiO}_{\text{2}}$ at 3000~K and 3500~K respectively, using a DZVP Gaussian basis set.}
\label{SiliconSilicaSquare}
\end{figure}

We turn now to more realistic problems such as those shown in FIG.~\ref{SiliconSilicaSquare}.
Although these simulations have been performed with only a single corrector step, they are still amazingly close to the BOMD results. It should be emphasized that even in liquid Si, which poses problems when using an ordinary Car-Parrinello scheme due to its metallicity, a single corrector step is sufficient. This establishes the efficiency of our method, since only a single preconditioned gradient calculation and no self-consistent iteration has to be performed. The possible acceleration, in comparison with regular BOMD calculations, depends crucially on the system studied. In these difficult cases a speed-up of two orders of magnitude compared to using the pure extrapolation scheme have been observed. For simpler problems an increase in efficiency of at least one order of magnitude can be expected.

\begin{figure}[tbp]
\includegraphics[width=8.6cm]{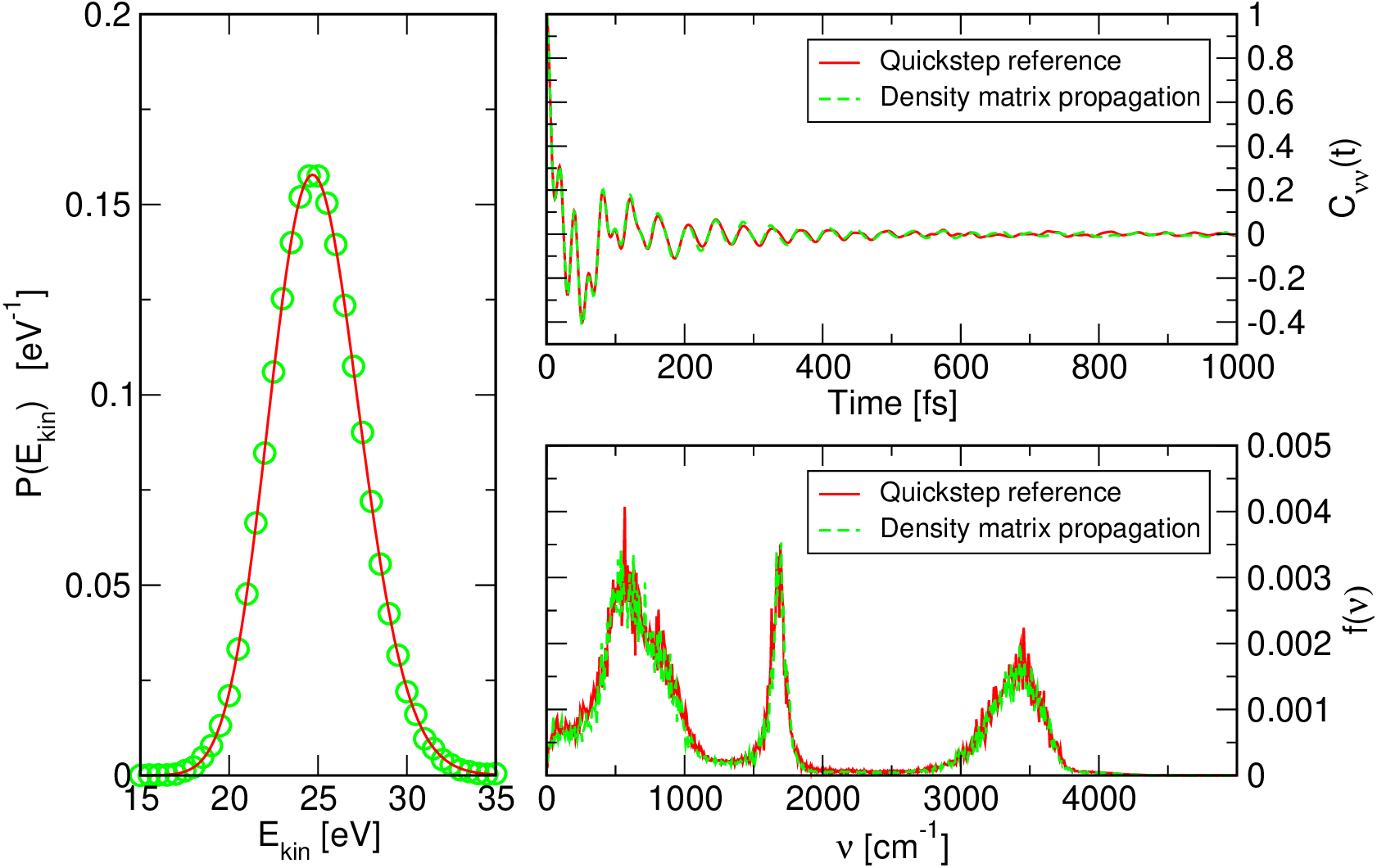}
\caption{The kinetic energy distribution calculated from a 1~ns trajectory
with a time step of 3.25~fs of metallic liquid $\text{Si}_{\text{64}}$ using
a DZVP Gaussian basis set and a density cutoff of 100~Ry (left panel).
Velocity autocorrelation function (upper right) and its Fourier transform
(lower right) of 32~Water at 325~K using a TZV2P Gaussian basis set, a
density cutoff of 280~Ry and the BLYP exchange-correlation functional. The
Langevin friction coefficients are $\protect\gamma_{L}=0$ and $\protect%
\gamma_{D} \sim 10^{-8}~\text{fs}^{-1}$.}
\label{KinEk}
\end{figure}

In FIG.~\ref{KinEk} we present results which prove that also dynamical properties can be evaluated with accuracy. This time we look at the velocity autocorrelation function and its Fourier transform at 325 K. The results are in good agreement with our reference calculations and are consistent with experiment and ab-initio all-electron calculations \cite{Krack2000}, showing that in spite of the stochastic nature of Eq.~(\ref{revLangevinEq}) dynamical properties can also be simulated. This implies, that also non-equilibrium processes and chemical reactions can be handled. In the same picture we verify that our 
assumptions are justified, and we are indeed performing a canonical sampling, by showing that the kinetic energy distribution is Maxwellian. To this effect, we have carried out a 64 atom liquid Si simulation for as long as 1~ns, to reduce the noise and to ensure a proper sampling of the kinetic energy distribution tails.

Because of space considerations we have reported here only a fraction of the systems studied. In all cases our method has proven to be accurate and the gain in speed has always been remarkable. Also structure relaxation via annealing and geometry optimization have been successfully performed. Contrary to CPMD and related methods at least BOMD integration time steps can be used. Thanks to this development it is now possible to perform ab-initio molecular dynamics on medium-sized systems up to a few nanoseconds, thus making a new class of problems accessible to ab-initio simulations. The computational scaling of the algorithm is in principle linear. However, an efficient parallel sparse matrix algebra has not been implemented yet, so that the sustained scaling is still $O(MN^{2})$, albeit with a much reduced prefactor.

In conclusion we wish to highlight that our ideas are of general interest, as our method is independent of the particular choice of the minimizer and propagation scheme, and can be further improved. In fact, during the submission process we became aware of a time-reversible extrapolation scheme \cite{NiklassonChallacombe2006}, which could be used as an alternative to ASPC. The possibility to apply the presented ideas with benefit to plane wave based CPMD is also to be underlined and will be presented elsewhere.

\begin{acknowledgments}
We would like to thank the whole CP2K team, in particular J. Hutter and J. VandeVondele for sharing their deep insights, and F. R. Krajewski and J. Kolafa for fruitful discussions. The generous allocation of computer time and support from CSCS Manno, the ICT Services of ETH Zurich and the DEISA consortium is acknowledged.
\end{acknowledgments}

\end{document}